\documentclass[11pt,aps,prd,reprint,groupedaddress,twocolumn,balancelastpage]{revtex4-1}

\usepackage{amssymb,amsmath,latexsym,graphics,graphicx,epsfig,multirow,comment,hyperref,appendix,verbatim,times,dcolumn,bm,threeparttable}

\usepackage{todonotes}
\newcommand{\mdw}[2][]{%
  \ifthenelse{\equal{#2}{}}%
  {\todo[color=red!40]{#1}}%
  {\todo[#1,color=red!40]{#2}}%
}

\usepackage{natbib}

\DeclareMathOperator{\sech}{sech}

\newcommand{\GeV}{\,\mathrm{GeV/c^2}}

\newcommand{\vmin}{{v_{\rm min}}}

\newcommand{\gsim}{\lower.7ex\hbox{$\;\stackrel{\textstyle>}{\sim}\;$}}
\newcommand{\lsim}{\lower.7ex\hbox{$\;\stackrel{\textstyle<}{\sim}\;$}}
\newcommand{\kms}{\text{ km s${}^{-1}$}}
\newcommand{\kpc}{\text{ kpc}}
\newcommand{\Gyr}{\text{ Gyr}}
\newcommand{\Msun}{\text{ M}_{\odot}}

\begin{document}

\title{The Dynamical Response of Dark Matter to Galaxy Evolution
  Affects Direct-Detection Experiments}

\author{Michael~S.~Petersen}
\email{mpete0@astro.umass.edu}
\author{Neal~Katz}
\author{Martin~D.~Weinberg}

\affiliation{University of Massachusetts at Amherst, 710 N. Pleasant St., Amherst, MA 01003}

\begin{abstract}
  Over a handful of rotation periods, dynamical processes in barred galaxies induce
  non-axisymmetric structure in dark matter halos.  Using n-body
  simulations of a Milky Way-like barred galaxy, we identify both a
  trapped dark-matter component, a \emph{shadow} bar, and a strong
  response wake in the dark-matter distribution that affects the
  predicted dark-matter detection rates for current experiments. The
  presence of a baryonic disk together with well-known dynamical
  processes (e.g. spiral structure and bar instabilities) increase the
  dark matter density in the disk plane.  We find that the magnitude
  of the combined stellar and shadow bar evolution, when isolated from
  the effect of the axisymmetric gravitational potential of the disk,
  accounts for $>$30\% of this overall increase in disk-plane density.
  This is significantly larger that of previously claimed deviations
  from the standard halo model.  The dark-matter density and kinematic
  wakes driven by the Milky Way bar increase the detectability of dark
  matter overall, especially for the experiments with higher
  $v_{min}$. These astrophysical features increase the detection rate by more
  than a factor of two when compared to the standard halo model and by
  a factor of ten for experiments with high minimum recoil energy thresholds. These same features increase (decrease) the annual
  modulation for low (high) minimum recoil energy experiments. We present
  physical arguments for why these dynamics are generic
  for barred galaxies such as the Milky Way rather than contingent on a specific galaxy model.
\end{abstract}

\pacs{95.35.+d, 98.62.Hr, 98.35.-a, 98.10.+z}

\maketitle

\section{\label{sec:intro}Introduction}

In the currently favored form of weakly interacting massive particle
(WIMP) theory (see e.g. \cite{jungman96,bertone05}), dark matter is
composed of a single particle with a mass in the range of 10 $\GeV~$, which a number of experiments are working to directly detect
\cite{aprile12,aalseth13,bernabei14,agnese15,angloher16,akerib16,agnes16,amole16,tan16,agnese16,undagoitia16}. Direct-detection
(DD) experiments seek to measure the weak nuclear recoils during
elastic scattering between dark-matter (DM) particles and the nuclei
of a target detector. The unambiguous detection of particle dark
matter would address fundamental questions about the nature of the
Universe, but despite considerable effort being focused on the direct
detection of dark matter, a verifiable signal remains elusive. Limits
on WIMP properties derived from these nondetections depend on poorly
constrained parameters from astrophysics \cite{mccabe10,mccabe11}.  The astrophysical uncertainties in the structure of the DM halo have
been recently implicated as a possible resolution for the disagreement
between experiments
with tentative detections (DAMA/LIBRA and CDMS-Si) and the null
results from experiments such as LUX and superCDMS \cite{mao13,pillepich14,bozorgnia16,kelso16,sloane16}.

\begin{table*} 
\centering
\begin{threeparttable}[b]
\caption{Halo Models\label{tab:simparams}.} 
\begin{tabular}{lccccc} Model Name & Designation\tnote{a}~~ & Radial Profile &
Dynamic? & Core? & Rotation?  \\ 
\hline \hline 
Standard Halo Model & SHM & isothermal & N & N & N \\
Pristine NFW & pNFW & NFW & N & N & N \\
Adiabatically Contracted NFW & acNFW & NFW & Y\tnote{b} & N & N\\
Fiducial Dynamical NFW & fdNFW & NFW & Y & N & N \\
Cored Dynamical NFW & cdNFW & NFW & Y & Y & N \\
Rotating Dynamical NFW & rdNFW & NFW & Y & N & Y \\
Cored Rotating Dynamical NFW & rcdNFW & NFW & Y & Y & Y \\
\hline
\end{tabular} 
\begin{tablenotes}
\item[a] Designations are used in Figures, Model Names are used in text.
\item[b] Idealized evolution; see text.
\end{tablenotes}
\end{threeparttable}
\end{table*}

Several simulation-based studies of Milky Way-like galaxies (e.g. a
multicomponent model featuring at a minimum a stellar disk and
responsive DM halo) have determined
velocity distributions for the DM halo that differ from the so-called standard halo model (SHM), finding that the spherical density and isotropic velocity distribution
assumptions underlying the interpretation of most DD experiments are unlikely to
be accurate owing to the presence of substructure in the halo
\cite{kuhlen10,purcell12,lisanti12}. Another class of
studies primarily focus on the difference between DM-only simulations and
simulations that include a stellar component
\cite{pillepich14,bozorgnia16,kelso16,sloane16}, finding largely the same results. However, little disagreement exists between
these studies regarding the expected response for DD experiments, and
the underlying dynamical causes have not been thoroughly investigated.

For example, these studies have been unable to reach a consensus on the
applicability of a Maxwell-Boltzmann (MB) distribution to describe the
DM
velocity distribution in the Milky Way (MW) near the Sun, and are roughly divided into groups
that claim a MB distribution does describe the tail of the DM velocity
distribution \cite{kelso16,bozorgnia16}, and those that find that the
tail is suppressed relative to a MB distribution
\cite{pillepich14,sloane16}.

In addition, the `dark disk', an axisymmetric, flattened DM feature
roughly on the size scale of the stellar disk observed in some
simulations, comprises an additional component for detection
\cite{read08,read09,bruch09,purcell09,ling10,pillepich14}, but its
existence continues to be debated. However, as we show in a previous work \cite{petersen15}, a dark
disk that mimics the appearance of the stellar disk is a natural
consequence of the presence of a stellar disk in a DM
halo, something that is obviously present in our own galaxy. The dark
disk effect may be enhanced further by the disruption of
satellites \cite{pillepich14}, which other studies contend may not be a generic result of
cosmological simulations \cite{kelso16}. This scenario is qualitatively
different from the dark disk
described in \cite{petersen15}. Other studies have claimed that the DM density at the Sun's
location should differ by less than 15\% from the average over a constant
density ellipsoidal shell using high resolution cosmological
simulations \cite{vogelsberger09} and that the density distribution is
only slightly positively skewed \cite{kamionkowski08}. Yet other
studies point out that many
open questions remain regarding the presence of substructure near the Sun owing
to either intact or destroyed subhalos
\cite{read08,read09,kuhlen12,lisanti12,ohare14}. In the face of these
conflicting claims, seeking fundamental effects from known Milky Way (MW)
causes is a prudent approach to illuminating the information that DM
halo models can provide for DD experiments.

Galaxies evolve structurally
through the interaction of the
baryonic matter in their disks with the DM in their halos mediated by
resonant gravitational torques.  The strongest evolution of this type
is likely to occur in barred galaxies (i.e. galaxies with prolate
stellar distributions in their central regions with lengths on the
order of the disk scale length).  The barred nature of the MW was first suggested in the 1960s as an interpretation of
observed gas kinematics \cite{devaucouleurs64}, and subsequently
confirmed through diverse observations in the ensuing half century
(see \cite{gerhard02} for a review). Recent observations have
indicated that the bar hosted by our MW galaxy may be significantly
longer than previously thought \cite{wegg15}. Although the MW bar is
known to have many consequences for observed astrophysical
quantities, the bar's effect on the DM distribution has not been
considered when characterizing the DM density and
velocity distribution function that determines detection rates for DD
experiments.

In this paper, we present the implications of non-axisymmetric DM
density and velocity distribution functions caused by the bar of the MW
for DD experiments.  We offer a qualitative analysis of recently published studies in an attempt to unify the seemingly disparate results. In a previous work \cite{petersen15}, we
demonstrated that particles in the DM halo will be trapped into a
shadow bar that resembles the stellar bar---in addition to forming a DM wake visible in both the density
and velocity structure of the dark matter halo at radii on the scale
of the stellar disk---the first such study
that attempts to isolate
the DM structure that results from interactions with the stellar
bar. The effect of the shadow bar is cumulative with the expected response of an equilibrium
galaxy DM halo to the presence of a stellar disk, resulting in a model
for the DM halo that does not resemble the SHM.  We will see that bar-driven galaxy evolution affects
both the DM density and the kinematics at the Earth's location.

Using simulations designed to study the mutual dynamical evolution
of the baryonic disk and DM halo for a Milky-Way-like galaxy, we
characterize the secular evolution of an initially exponential stellar disk
and spherically symmetric dark matter halo. We do
not consider any satellite debris or stellar streams at the solar
circle \cite{freese04,savage06}, although these may be present.
Rather, we detail significant differences from the SHM
due to the stellar bar of the MW. Similar to
previous studies \cite{mao13}, we find that realistic DM distributions
in galactic halos can dramatically increase the predicted detection rates
for high $v_{min}$ experiments.  Moreover, the effects of long-term
evolution in a barred galaxy further increases the tension between
heavy and light nuclei experiments \cite{frandsen13}. We demonstrate
key regimes in which experiments can use the DM halo structure resulting
from the MW bar to their advantage.  Conversely, \cite{pillepich14}
report an improvement in the tension between the heavy and light
nuclei experiments if the detection signal were dominated by a DM debris
disk from merger events, which has a sharply decreasing velocity tail.
It is possible, of course, that the MW also has a DM debris disk from a
merger event.  This underscores the importance of the actual MW
evolutionary history to DM detection predictions and motivates further
detailed study.

This paper is organized as follows.  In section~\ref{sec:fiducial}, we
provide the relevant details about the simulations used for this analysis,
including a comparison of the simulations to the MW in
section~\ref{subsec:milkywaycompare}. We then describe the results in
section~\ref{sec:results}, beginning with the density and kinematic
features of the simulated galaxy in section~\ref{subsec:features} before
detailing the calculation of detection rates in
section~\ref{subsec:detectionrates}. We compare to previous
findings in section~\ref{subsec:litmod} (including both the SHM and
empirical models), then explore the effect of our results for detection rates in DD
experiments (sections~\ref{subsec:experiments} and \ref{subsec:annualmod}).
Section~\ref{sec:conclusion} provides a broad overview of our results
and prospects for future work.

\section{Methods} 
\label{sec:fiducial} 

\subsection{Simulations} 
\label{subsec:methods}

\begin{table*} 
\centering
\caption{List of Milky Way Disk Scale Lengths in the literature\label{tab:scales}.} 
\begin{tabular}{lc} Method & Scale Length (kpc) \\ 
\hline \hline 
Asymptotic Giant Branch Stars \cite{nikolaev97} & $4.00\pm0.55$\\
COBE/DIRBE \cite{bissantz02} & 2.1 \\
G-dwarfs ($\alpha$-old) \cite{bovy12} &2.01 $\pm$ 0.05 \\
G-dwarfs ($\alpha$-young) \cite{bovy12} &3.6 $\pm$ 0.22 \\
G-dwarfs (mass-weighted) \cite{bovy13} & 2.15 $\pm$ 0.14 \\
\end{tabular} 
\end{table*}

The n-body simulations analyzed here are presented in
\cite{petersen15}. We summarize the initial conditions for their
relevance to the results and refer the interested reader to
\cite{petersen15} for details of the simulation methodology and
dynamical interpretations. We list the simulations used in this paper
in Table~\ref{tab:simparams}.

We represent the axisymmetric disk density profile by an exponential
radial distribution with an initially isothermal $\sech^2$ vertical
distribution, consistent with observations of the MW
\cite{bovy13}. The DM halo is a fully self-consistent,
cosmologically-motivated DM halo \cite[][NFW]{navarro97} with $c =
R_{\rm vir}/r_s \approx15$ where $r_s$ is the scale radius, and
$R_{\rm vir}$ is the virial radius. The functional form of the NFW
profile is given by
\begin{equation}
\rho(r) \propto \frac{r_s^3}{(r+r_{\rm core})(1+r_s)^2}
\label{eq:nfw}
\end{equation} 
Observations of the central density profile in the MW are consistent
with either a pure NFW profile or a cored NFW profile \cite{bovy13}.
The latter choice is motivated both by observational data and
dynamical theory: a cored halo is more likely to be unstable to bar
formation.  We therefore test exampless of both models by selecting $r_{\rm core} =
0.0$ or $0.02$. We call the model with $r_{\rm core}=0.0$ the {\it
  fiducial dynamical NFW} model, and use it as the primary model
throughout our work. The $r_{\rm core}=0.02$ model is called the {\it
  cored NFW} model. In practice, the cored halo model increases the relative disk
density to halo density near the center of the simulation, while
causing a variation of 20\% at the approximate solar radius. We
construct these initial halos without rotation, but acknowledge that
true DM halos are expected to have some net rotation \cite{bullock01};
we present two additional models with modest rotation to
probe any possible effects. The $r_{\rm core}=0.0$ and $r_{\rm
  core}=0.02$ rotating models are called the {\it rotating NFW} and
{\it cored rotating NFW} model, respectively.

Our simulations employ $N_{\rm disk}=10^6$ and $N_{\rm halo}=10^7$,
disk and halo particles, respectively.  These values ensure there is enough
phase-space coverage to model resonant torques and to
resolve collective features such as stellar bars and spiral arms. The
disk particles have equal mass and the halo-particle masses are assigned
to satisfy both the NFW density requirement with a steeper number density distribution, $n(r) \propto r^{-2.5}$.  Relative to an
equal-mass assignment, this improves the resolution of the mass and length scales
in the gravitational potential of the DM halo by a factor of
approximately 100 in the vicinity of the stellar disk, i.e. it is
equivalent to the resolution of a $N_{\rm halo}=10^9$ model.

A DM halo in dynamical equilibrium will respond to the growth of a
baryonic disk through dissipation.  This slow-growth process is often
modeled in the adiabatic limit and is called `adiabatic contraction'.
It causes the halo density profile to become mildly oblate in response
to the disk potential.  To test the importance of this process, we
additionally draw on the results of a simulation presented in
\cite{petersen15} that artificially freezes the stellar disk profile
while the DM halo self-consistently evolves.  While not strictly an
adiabatic process, we refer to this as the \emph{adiabatically
  contracted NFW}
model.

We also compare the dynamically evolved models listed above to the static
{\it pristine NFW} model given by eq~\ref{eq:nfw} with $r_{\rm
  core}=0.0$, as well as the {\it standard halo} model (SHM).

\subsection{Calibrating to the Milky Way}
\label{subsec:milkywaycompare}

\subsubsection{Dynamical Units} 
\label{subsubsec:dynunits}

\begin{table*} 
\centering
\begin{threeparttable}[b]
\caption{List of MW Bar parameters in literature\label{tab:barparams}.} 
\begin{tabular}{lcc} Method & Bar Length (kpc) & Bar Angle \\ 
\hline \hline 
Asymptotic Giant Branch Stars \cite{nikolaev97} & 3.3 $\pm$0.1 & 24$^\circ\pm$2$^\circ$ \\
OH/IR Stars \cite{sevenster99} & $<3.5$\tnote{a} & -- \\
near-infrared photometry \cite{hammersley00} & 4.0 & $43^\circ\pm7^\circ$\\
Local stellar velocities \cite{dehnen00} & $<$5.3\tnote{a} & $20^\circ-45^\circ$ \\
COBE/DIRBE  \cite{bissantz02} & 3.5 & $20^\circ-25^\circ$ \\
near-infrared photometry  \cite{babusiaux05}& $$2.5 & 22$^\circ\pm 5.5^\circ$ \\
Red Clump Giants (UKIDSS) \cite{cabrera08} & 4.5 & $42.44^\circ \pm$ 2.14$^\circ$ \\
Methanol Masers \cite{green11} & $<3.3$\tnote{a} & 45$^\circ$  \\
Red Clump Giants (compilation) \cite{wegg15} & $5.0 \pm 0.2$&
                                                                   $28^\circ-33^\circ$\\ 
\hline
\end{tabular} 
\begin{tablenotes}
\item[a] Denotes a measurement of corotation, considered to be an
  upper limit for the bar length.
\end{tablenotes}
\end{threeparttable}
\end{table*}

\begin{table*} 
\centering
\begin{threeparttable}[b]
\caption{Physical versus Simulation Parameters for the Milky
  Way\label{tab:mwparams}.}
\begin{tabular}{lcc} Quantity & MW Value & Simulation Value \\ 
\hline \hline 
Scale Length, $R_d$ & 2.01-4.00~\kpc~(see Table \ref{tab:scales}) & 3 kpc \\ 
$R_\odot$ Scale Height &0.37 $\pm$0.06 kpc \cite{bovy13}& 0.3 kpc \\ 
Disk Mass (Stellar) & $4.6 \pm 0.3 ({\rm ran.}) \pm 1.5 ({\rm syst.})\times10^{10}~\Msun$ \cite{bovy13} & $3.25\times10^{10}~\Msun$\\ 
Halo Mass & $1.6 \times10^{12}~\Msun$ \cite{boylan13}& $1.6 \times10^{12} ~\Msun$\\
Virial Radius & $304 \pm 45$~\kpc~\cite{garrisonkimmel14} & 300 kpc\\
$R_\odot/R_d$\tnote{a} &  2.08-4.13 (see
                          Table \ref{tab:scales}) & 2.08-4.13 (see Section \ref{subsubsec:dynunits})\\
$R_\odot/R_{\rm bar}$\tnote{a}&  1.57-3.32~\cite{reid14,chatzopoulos15} (see
                          Table \ref{tab:barparams}) & 1.57-3.32 (see Section
                                                       \ref{subsubsec:dynunits})\\
$R_\odot$ Circular Velocity &  $218 \pm 10 ~\kms$~ \cite{bovy12} & 218~\kms\\
\end{tabular} 
\begin{tablenotes}
\item[a] Using $R_\odot$=8.3~\kpc~\cite{reid14,chatzopoulos15} 
\end{tablenotes}
\end{threeparttable}
\end{table*}

We scale the dynamical units of the simulations to the
mass of the MW halo without attempting to tune the initial conditions
to produce a model that more closely matches the details of the MW
(e.g. its rotation curve, bar length, and bar amplitude). We plan to
more closely mimic the MW in future simulations. We select a snapshot
of the simulation after initial bar formation ($T=1$ Gyr) and a
subsequent `secular evolution time' $\Delta T_{\rm se}=
3$ Gyr, defined as the time after the bar has formed, during which the
bar strengthens and grows in length as a result of continued angular
momentum transfer by secular processes (see \cite{petersen15}). In general, the results are
qualitatively similar for all outputs after bar formation. We
discuss possible variations owing to the time selection where relevant.

To better compare the MW with the
simulation, we may choose to scale the Galactic radius of the solar
position to the disk scale length, to the bar length, or to something
in between.  The first scaling is fraught with astrophysical
uncertainties, such as the variation of disk scale length with metallicity.
This induces a dependence on the age of the stellar population used to
estimate the disk scale length. In Table \ref{tab:scales}, we list some
literature measurements of the disk scale length.  Comparing to our
simulation, we find that the Sun could be located anywhere between two
and four disk scale lengths.  The uncertain location of the Sun in the
phase-space of the halo has been previously described as a large
source of uncertainty \cite{mao13}. We, therefore, report a range of
results that correspond to the uncertainty for the location of the Sun
in this model. As noted in \cite{petersen15}, further study of the MW
bar history will reduce uncertainties related to scaling simulations
to the MW.

Scaling to the length of the bar better represents our goal
of studying the influence of the bar on the DM distribution at the
solar position.  Nevertheless, calibration to the bar is also uncertain owing
to the diversity of parameter measurements for the MW bar in the literature. In Table
\ref {tab:barparams}, we list bar parameters measured for the
MW. Using this scaling, the Sun is located between 1.57 and 3.32 bar
radii. We choose a nominal scaling of 2 bar radii for the Sun as a
compromise between measurements of the disk scale length and bar
radii. Additionally, \cite{wegg15} presents a bar mass in the range of
$1.1-1.81\times10^{10}~\Msun$, or $0.24-0.39M_{\rm disk}$ (using the
scaling from \cite{bovy13}). This broadly agrees with the bar mass in
the simulation at $\Delta T_{\rm se}=3$, which we find to be
$0.35M_{\rm disk}$.

Since the Sun is measured to be only 25 pc above the disk midplane
\cite{juric08}, and this is smaller than the resolution scale of our
simulation, we will consider the Sun to be in-plane for the purposes
of our calculations here. In practice, this introduces errors below
the 1\% level. Throughout the paper, in-plane refers to
$|z|<1~\kpc$. As in previous simulation-based studies
\cite{kuhlen10,purcell12,pillepich14}, we define a region of interest
around the solar neighborhood from which to draw velocity
samples. To achieve an accurate velocity distribution with the desired
spatial sampling, we create wedges 1 kpc in radius, 2 kpc in height,
and $\frac{\pi}{7}$ in azimuth. In addition, we sum 20 phase space outputs
(total $\delta
T= 0.08 \Gyr$) near $\Delta T_{\rm se}=3$ in a frame of reference rotating with the stellar bar,
to decrease the noise further. Each bin has $>10^{5}$ particles.

We caution that the scalings presented in this paper are tied to
the virial mass of the Milky Way DM halo, with a linear scaling in
density, but a much more complex and poorly understood effect on the
velocity structure. We choose a virial halo mass of $1.6\times
10^{12}~\Msun$, as determined from the motion of the MW satellite Leo
I \cite{boylan13}. Local stellar kinematics imply a halo mass of $8
\times 10^{11}~\Msun$ \cite{bovy12} and suggests a factor of two
uncertainty in this calibration. In addition, the rotation curve in
our model deviates from the estimates of the MW rotation curve in
\cite{bovy13}; the rotation curve in our simulation is slowly rising
inside of three disk scale lengths rather than flat. We cannot comment
quantitatively on the importance of the relative disk-to-halo
potential contribution in the inner galaxy, a quantity that is poorly
constrained in the MW as well \cite{bovy13}.

\subsubsection{Velocity Definitions}
\label{sec:veldef}

The velocity of the Earth in the MW relative to the galaxy's inertial
frame is the sum of three terms
\begin{equation}
\label{eq:evel}
\vec{v}_e(t) = \vec{v}_{\rm LSR} + \vec{v}_\odot + \vec{v}_\oplus(t).
\end{equation}
where $\vec{v}_{\rm LSR} $ is the \emph{local standard
  of rest} (LSR), $\vec{v}_\odot$ is the peculiar motion of the Sun,
and $\vec{v}_\oplus(t)$ is the relative motion of the Earth. It is traditional to define the LSR as the mean motion of stars in the neighborhood of
the Sun on a hypothetical orbit about the center of the Galaxy.  This
hypothetical orbit need may not circular, although circularity is
often assumed.  We define the three velocity directions $U,~V,~W$ in the LSR frame as
follows: $U$ points toward the Sun from the Galactic Center, $V$
points in the direction of Galactic rotation, and $W$ points perpendicular to the
Galactic disk.  The first velocity in equation (\ref{eq:evel}) is the
velocity of the LSR relative to the Galactic Center.  We adopt
$\vec{v}_{\rm LSR} = (0,218\pm6,0) $\kms \cite{bovy12}.  The second
term is the motion of the Sun relative to the LSR, the peculiar
velocity, defined as $\vec{v}_\odot =
(11.1^{+0.69}_{-0.75},12.24^{+0.47}_{-0.47},7.25^{+0.37}_{-0.30})
\kms$ \cite{schonrich10}, though somewhat larger values of
$U_{\odot}=14$ \kms \cite{schonrich12} and $V_\odot =
23.9^{+5.1}_{-0.5} $\kms \cite{bovy12} have been reported. The third
term is the motion of the Earth in orbit around the Sun, for which we
follow \cite{lewin96}. For the purposes of this study, we will
consider only the velocity maxima and minima for the alignment and
anti-alignment, respectively, of the Earth's velocity with the LSR
motion.  These epochs provide the largest kinetic energy difference
and occur on approximately June 1, $V_\oplus = (0,27.79,0)$ \kms, and
on December 1, $V_\oplus = (0,-27.79,0)$ \kms, using the standard
speed for the Earth of 27.79~\kms. This simple parameterization of the Earth's
velocity relative to the Sun avoids the discrepancy in \cite{lewin96}
pointed out by \cite{lee13,mccabe14}.  

We scale the simulations to select $v_{\rm LSR}$ as the azimuthal
velocity at the solar radius, as chosen in section
\ref{subsubsec:dynunits}. The scaling to the $v_{\rm LSR}$ (as well as the corresponding peculiar motions of the Sun relative to LSR) comprises the largest
uncertainty in our comparison, but we emphasize that the relative
importance of the shadow bar for the direct detection of DM
remains unchanged.

\begin{figure*} 
\centering 
\includegraphics[width=5.5in]{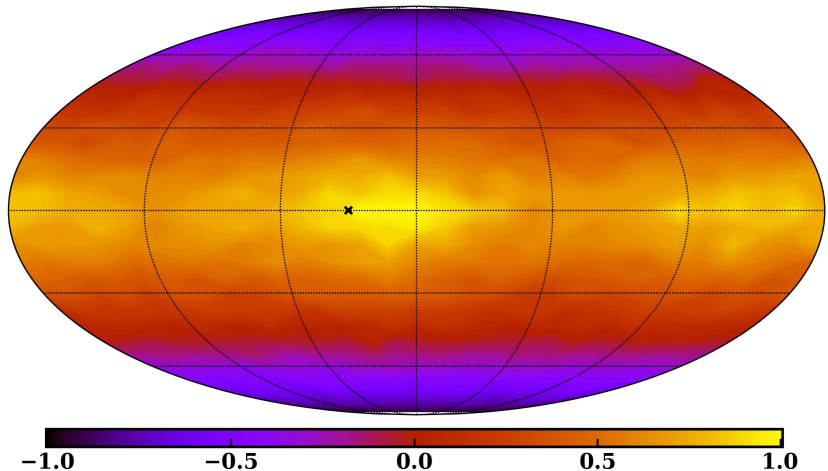} 
\caption{Mollweide projection of the relative DM density deviation at the solar
  radius to the mean DM density at the same radius for the fdNFW model. The coordinate
  system is oriented such that the bar angle is
  (0$^\circ$,180$^\circ$). The approximate position of the Sun is
  marked with an `{\bf x}'. The flattening of the halo is clearly seen
  as a decrease in the density at the poles. The effect of the
  bar is seen as peaks at approximately
  (-15$^\circ$,165$^\circ$)\label{fig:soldarkdens}.}
\end{figure*}

\subsubsection{Summary of Key Differences}

The fiducial dynamical NFW model results in a barred galaxy that has many 
properties similar to the MW. However, we identify two potentially important
differences:
\begin{enumerate}
\item The fiducial model does not have a flat rotation curve at the solar
  circle in contrast to observations \cite{bovy13}, and thus the
  tuning of velocity in the simulation to that of the MW may have some
  systematic errors. The choice of $v_{\rm LSR}$ affects the width of
  the calculated speed distribution through the dispersion.
\item The ratio of the length of the bar to the disk scale length may
  suggest a different (i.e. triggered) origin for the MW bar, possibly
  from an orbiting satellite such as the Sagittarius dwarf
  \cite{purcell11} or the Large Magellanic Cloud whereas our simulation
  forms a bar in isolation.
\end{enumerate}
We comment on the possible effects of these differences at relevant
points throughout the paper, and again stress that the model has not
been specifically tuned to the MW, but should rather be considered
MW-like. Table \ref{tab:mwparams} provides a concise comparison of
measured MW parameters to the
simulation parameters, valid for all NFW-derivative halo models.

\section{Results} \label{sec:results}

We begin this section by reporting the salient differences between
static and dynamically evolving galaxy models that affect the DD rate.
We describe the DM density and velocity variations in response to the
bar in section \ref{subsec:features}. We compute the detection rates
in section \ref{subsec:detectionrates}. In this section we restrict
our analysis to the fiducial dynamical NFW model, comparing to other
models in sections~\ref{subsubsec:initnfw},
\ref{subsubsec:ohalos}, and \ref{subsubsec:shm}.

\subsection{Dark Matter Distribution Features} 
\label{subsec:features}

We begin with a discussion of the self-consistent response of the DM
halo to a bar-unstable disk. There are two clear
deviations from a spherical distribution: flattening (Section
\ref{subsubsec:dmdens}), and non-axisymmetric contributions due to the
bar (Section \ref{subsubsec:wake}). We then analyze the velocities and
speed distribution in Section \ref{subsubsec:dmvels}.

\subsubsection{The Dark Disk}\label{subsubsec:dmdens}

\begin{figure*} \centering 
\includegraphics[width=5.5in]{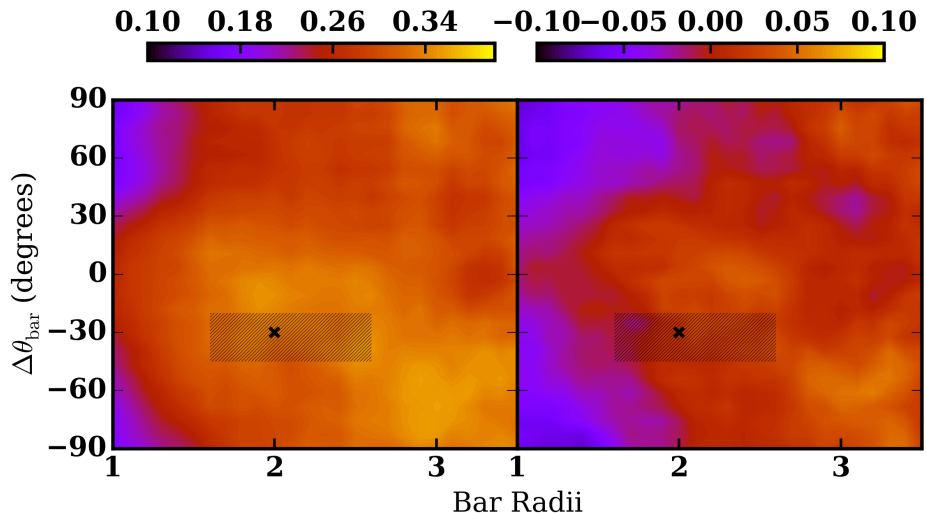} 
\caption{In-plane relative DM density as a function of bar radius and
  bar angle for the fdNFW model. Left panel: the simulation at $T=4 \Gyr$ versus the
  pristine NFW model. Right panel: the simulation at $T=4
  \Gyr$ versus an adiabatically contracted model. The best choice solar
  position is marked with an `x'. The possible solar locations
  consistent with astronomical uncertainties are denoted by the
  hatched region. Both panels show similar features, including a
  quadrupole disturbance owing to the bar that appears as a density
  enhancement trailing the bar. The patchiness in the relative
  density determinations owe to the self-consistent evolution
  (see \cite{petersen15} for further discussion). \label{fig:rthetadens}}
\end{figure*}

As a first characterization of the halo structure, we compute the
ellipsoidal axes by diagonalizing the moment of inertia tensor as in
\cite{allgood06}. Similar to the findings of \cite{kuhlen07}, we find
that the halo becomes flattened owing to the presence of the disk with
$\left(c/a=0.5\right)_{\rm fdNFW}$ at the chosen solar radius, where $c$ and $a$ are the minor and major
ellipsoidal axes, respectively. We find that
$\left(c/a=0.6\right)_{\rm acNFW}$ at the chosen solar radius for the adiabatically contracted
NFW model.  Fitting a disc and
NFW halo model potential to the vertical
structure of halo giant stars in the MW suggests $c/a=0.8$ at the solar
circle \cite{bienayme14,piffl14}, a smaller deviation from spherical than our
findings. However, this ratio is poorly constrained by
presently available data. The apparent disagreement
may reflect the complexity of modeling the DM distribution from
stellar data more than a problem with our models.  For example, the
halo stars at large distances from the disk are likely the result of
hierarchical formation and satellite accretion and are unlikely to be
affected by the environmental processes that affect DM near the disk
in our simulation. 

Figure~\ref{fig:soldarkdens} illustrates the deviation from a
spherical distribution by showing a Mollweide projection of the relative
density on a sphere at the solar radius:
$(\rho - \langle\rho\rangle)/\langle\rho\rangle$. 
The approximate position of the Sun is marked, showing that the Sun
resides in a strongly over-dense region in our simulation relative to
the spherical average.  Two effects are clearly at play in causing the
density of the DM halo to deviate from spherical. The first is the
compression towards the disk plane, which is clearly seen
as a gradient from low latitude to high latitude. The second,
variations in longitude (non-axisymmetric structure), will be
discussed in the following section.

The compression of the halo to an oblate figure is caused by two
independent dynamical effects. The first, adiabatic contraction, is a
response of the spherical halo to the potential of the embedded
stellar disk. However, as noted above, $(c/a)_{\rm fdNFW} <
(c/a)_{\rm acNFW}$, i.e., the fiducial dynamical NFW model is
more oblate than the adiabatically contracted NFW model. This extra
contraction is caused by the bar, which torques the halo through secular
resonant interactions (see \cite{petersen15} for further dynamical
details). $\left(c/a\right)_{\rm fdNFW}$ decreases as $\Delta T_{\rm se}$ increases, suggesting
that the in-plane density may not have been as large in the past.

We refer to the enhanced (in-plane) contraction as the \emph{dark
  disk} owing to its phase space resemblence to the stellar disk, while
noting that previous works have used this term to refer to shredded
satellites that contribute DM in a kinematic disk-like structure
\cite{read08,bruch09,read09,purcell09,pillepich14}.  The similarity of
the DM distribution to the stellar distribution at corresponding radii
is discussed in section~\ref{subsubsec:dmvels} and extensively in
\cite{petersen15}. As discussed in \cite{petersen15}, the primary driver of large-scale aspherical structure in the DM halo is the combination of the stellar disk and bar. We do not find any evidence for the claim that the presence of baryons in a simulation will make the halo more spherical \cite{kelso16}.

\subsubsection{The Shadow Bar and Density Wake}
\label{subsubsec:wake}

In addition to the dark disk creating an axisymmetric overdensity,
the stellar and dark-matter shadow bar create non-axisymmetric density variations
that correspond to a global quadrupole. This response of the DM halo
to the stellar bar results in a collisionless wake; this wake appears
as a diffuse $m=2$ spiral (see Figure 6 of \cite{petersen15} for details).  The effect of this DM feature is readily
seen in Figure \ref{fig:rthetadens}, which plots the in-plane relative
DM density as a function of bar radius and bar angle. When comparing
the fiducial NFW model to the pristine NFW distribution (left
panel), we see a clear density enhancement at a $>$15\% level
everywhere, peaking at $>$40\% lagging just behind the bar at two bar
radii. At the approximate solar location, we find a 35$^{+5}_{-3}$\%
enhancement relative to a spherical distribution.

When we compare to the adiabatically contracted NFW model to isolate the effects
of the stellar and shadow bar (right panel), we find that the
fiducial NFW model exhibits an over density along the bar
major axis relative to the minor axis of approximately 15\% at $T=4
\Gyr$ at the solar circle, corresponding to $>30$\% of the total
effect when compared to the difference between the fiducial NFW and
pristine NFW
models. The fiducial dynamical NFW model has an average
of 10\% (30\%) greater density everywhere when compared to the adiabatically
contracted NFW model (pristine NFW model). The fiducial model has a lower
azimuthally-averaged density within two bar radii, caused by the
transport of angular momentum from the stellar disk, making the
DM
orbits gain in net angular momentum and thus experience some radial
expansion.

\subsubsection{Dark Matter Kinematic Wake}\label{subsubsec:dmvels} 

\begin{figure} 
\centering 
\includegraphics[width=3.5in]{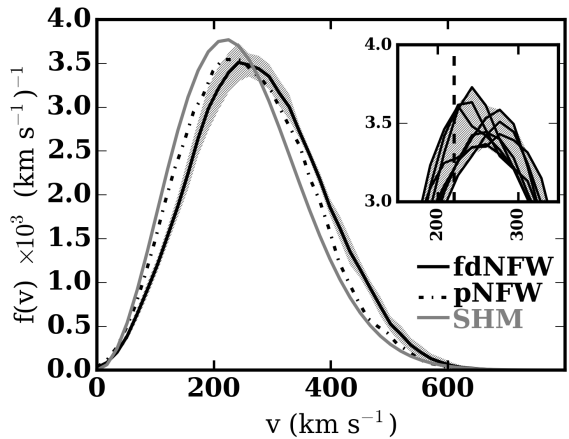} 
\caption{Speed distribution at the solar position in three different
  halo models. The
hatched region around the fdNFW line indicates the extent of the possible solar
  locations in the simulation. The pNFW model is plotted as a dot-dashed
  line. The SHM is plotted as a dashed
  line. Inset: zoom-in on the peak of the speed distribution, with the
  extent of the solar position uncertainty indicated as a shaded
  band. Thin lines represent individual realizations of the region of
  interest used to calculate the solar position speed
  distribution. $|v|=220~\kms$, the peak of the SHM, is marked
  as a vertical dashed line. Note that peaks for individual
  realizations range between 230 and 280$~\kms$. \label{fig:solvel}}
\end{figure}

\begin{figure} 
\centering 
\includegraphics[width=3.5in]{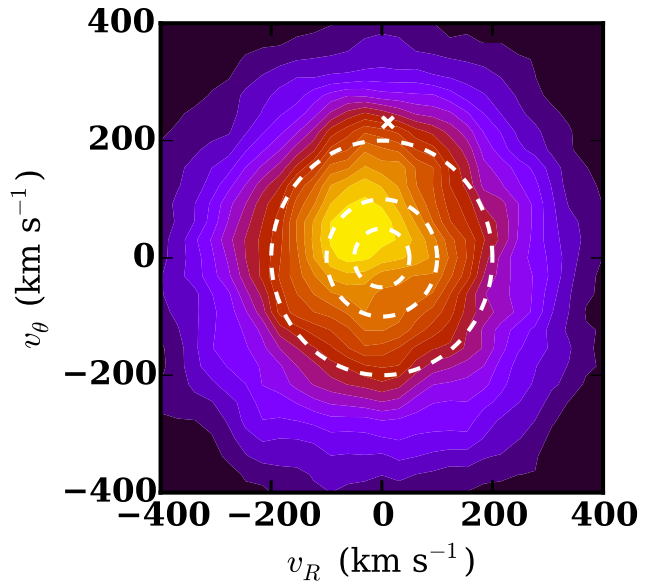} 
\caption{Radial ($v_r$) versus tangential ($v_\theta$) velocities in
  galactocentric coordinates at the solar position for the fdNFW model. To illustrate the
  deviation from an isotropic distribution, we plot circles with $|v|
  = 50,~100,~200 \kms$. The velocity of the Sun in $v_r-v_\theta$
  space is marked with a white 'x'.\label{fig:correlatedvel}}
\end{figure}

In Figure~\ref{fig:solvel}, we plot the speed distribution at the
solar circle. We choose the solar circle as nine regions of interest
centered at each combination of $R=[1.6,2.0,2.4]$ bar radii and
$\Delta\theta_{\rm bar} = [20^\circ,~30^\circ~40^\circ]$. We plot the
speed distribution for the SHM,
which is a MB distribution centered at 220 \kms, as a dashed
line for comparison. The peak of the speed distribution shifts upward,
and is
now between 230 and 280$~\kms$ with more populated tails than in
the standard MB distribution. The shift in the peak relative to the SHM is caused by a
non-isotropic velocity structure in the DM halo, which is evident in
Figure~\ref{fig:correlatedvel}. 


The shape of the
distribution depends on both the initial phase-space distribution
and the galaxy's evolutionary history, so we can not provide a generic
parametrization at this time. The magnitude of the wake increases with
$\Delta T_{\rm se}$, meaning that an older bar with more time to
transfer angular momentum to the halo will enhance the azimuthal
velocity of orbits in the halo.

Similar to \cite{pillepich14}, we opt not to fit a MB distribution to
the peak of the speed distribution. As noted by \cite{mao13}, the
MB distribution does not provide a good fit to the speed
distribution.  We demonstrate in section~\ref{subsec:detectionrates}
that the underlying reason a 
MB distribution is a poor descriptor for our DM velocity distribution owes to a combination of adiabatic contraction and the
stellar+shadow bar. The underlying distribution may not be well
described by a single fitting-function
parametrization dependent upon escape velocity (e.g. \cite{mao13}).  

In Figure~\ref{fig:correlatedvel}, we plot the distribution of the
radial ($v_r$) versus azimuthal ($v_\theta$) velocity components in
galactocentric coordinates. The shift in the peak of the azimuthal velocity distribution, $\delta v_\theta=+50~\kms
$ shows that the dark disk has gained net
rotation. In addition, the peak in radial velocity has been
decreased owing to the DM wake induced by the bar ($\delta v_r=-30~\kms$).  A similar analysis
performed on the adiabatically contracted NFW model yields a nearly isotropic
distribution.  Thus, the bias of the velocity distribution to higher
tangential velocities and lower radial velocities solely owes to the non-axisymmetric evolution of the disk, i.e. the bar, without a
net gain in angular momentum.

Despite concerns that the dark disk could inhibit direct detection of
DM \cite{billard13} by causing 10-50\% of the DM at the solar radius
to co-rotate (consistent with our findings), we find
that the formation mechanism of the dark disk increases the tails of
the velocity distribution and, thus, increases the fraction of particles with 
velocities greater than prospective values of $v_{\rm min}$. The speed distribution is
shifted to significantly higher velocities, with the tail falling more
steeply than that of the SHM, similar to the findings of several studies 
\cite{mao13,pillepich14,sloane16}. The implications of the tails for DD experiments are
discussed in section \ref{subsec:detectionrates}.

In summary, we find that the
stellar+shadow bar causes the halo in our simulation to deviate from
the standard halo model in three important ways: (1) the presence of
the stellar disk potential causes the halo to contract toward the
plane, producing an oblate spheroid; (2) the stellar+shadow bar causes
a density enhancement along the bar axis; and (3) the stellar+shadow
bar causes a further contraction toward the plane and a
non-isotropic velocity distribution by transferring angular momentum
to the dark disk. Future simulations matched in detail to the MW will be able to
provide a more nuanced understanding of the shape and structure of the
speed distribution.

\begin{figure} 
\centering 
\includegraphics[width=3.5in]{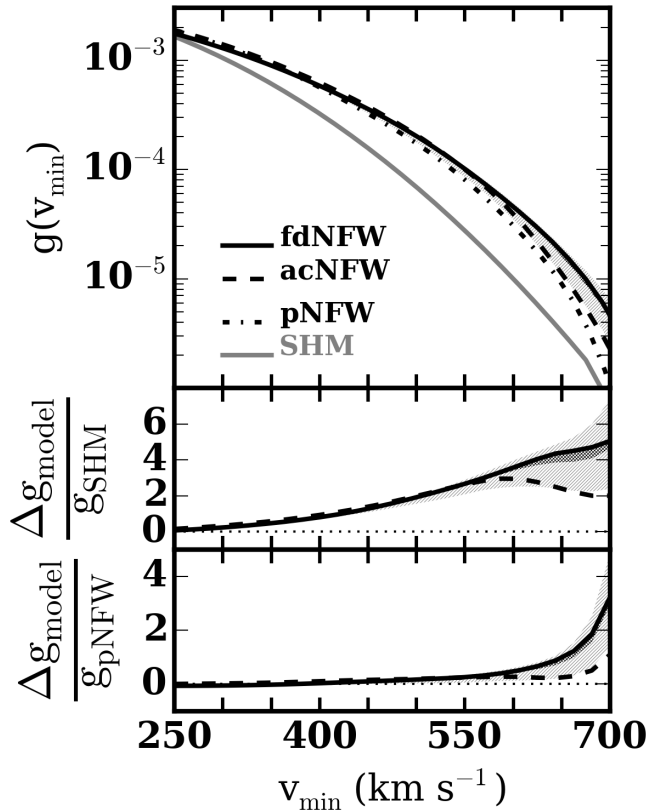} 
\caption{Upper panel: $g(\vmin)$ as a function of $\vmin$ for the
  fiducial dynamical NFW model. The best fit solar location is shown as a solid black
  line. The uncertainties due to the solar position are shown; the
  combination of radial and azimuthal uncertainty is lightly shaded,
  while the azimuthal uncertainty alone is darkly
  shaded. The pristine NFW distribution and adiabatically contracted
  NFW
  distributions are shown as the dot-dashed and dashed black lines,
  respectively. The standard halo model is shown as a solid gray
  line. Middle panel: comparison of the empirical simulation results
  to the SHM. The solid black line shows the relative
  value of $g(\vmin)$ ($(g(\vmin)_{\rm NFW} - g(\vmin)_{\rm
    SHM})/g(\vmin)_{\rm SHM}$) for the most likely solar position in
  the empirical NFW halo to $g(\vmin)$ for the standard halo
  model. The lightly-shaded region shows the uncertainty only due to the
  radial and azimuthal uncertainty and the darkly-shaded region shows
  the uncertainty due to the azimuthal uncertainty. The dashed black
  line shows the same quantity for the adiabatically contracted NFW
  model. Bottom panel: comparison of the empirical simulation results
  to the pristine NFW distribution ($(g(\vmin)_{\rm NFW} -
  g(\vmin)_{\rm pNFW})/g(\vmin)_{\rm pNFW}$). The lines are the same
  as in the upper and middle panels.\label{fig:geevee}}
\end{figure}

\subsection{Dark Matter Detection Rates} 
\label{subsec:detectionrates}

In this subsection, we present computations for the DD rates, as well
as a physical justification for the observed phenomena. We first discuss
the detection of DM in general, then move to the fiducial dynamical
NFW model, the
adiabatically contracted NFW model, and the pristine NFW model (sections
\ref{subsubsec:fidmodel} and 
\ref{subsubsec:initnfw}, respectively). We also qualitatively discuss
the results of other halo models presented in \cite{petersen15}
(section \ref{subsubsec:ohalos}). Taken together, these sections
implicate the self-consistent dynamical evolution in the fiducial
model as the driver of the observed variation in expected detection
rates, the principal finding of this work.

Following other studies that compute the magnitude of these effects for
DD experiments (e.g. \cite{kuhlen10,purcell12}), we calculate
differential event rates, in counts per day per unit nucleus mass per
unit exposure time per unit velocity (cpd/kg/(\kms)), as a function of
the minimum velocity ($\vmin$) using the new density and speed
distributions obtained from the simulations:
\begin{equation}
\frac{dR}{d\vmin}\left(\vmin\right) =
\frac{\sigma_\chi}{2\mu m_\chi}\rho_0 g(v_{\rm min})
\label{eq:detect}
\end{equation}
where $\sigma_\chi$ is the spin-independent WIMP cross-section for scattering on a
proton, $\rho_0$ is the WIMP density in the
solar neighborhood, $m_\chi$ is the WIMP mass, $\mu=(m_N
m_\chi)/(m_N+m_\chi)$ is the WIMP-nucleus reduced mass, and the
quantity $g(v_{\rm min})$ is the integral in velocity space of the
speed distribution divided by the WIMP speed,
\begin{equation}
g(v_{\rm min}) = \int_{v_{\rm min}}^\infty \frac{f(v)}{v}dv.
\label{eq:geevee}
\end{equation} 
The threshold speed, $\vmin$, can be translated to the nuclear recoil
energy $E_R$ via the relation $v_{\rm min} =
\sqrt{\frac{E_Rm_N}{2\mu^2}}$ for calculating specific
experimental detection rates.  In the interest of exploring the
astrophysical variations, we ignore the nuclear form factor and
dependence on recoil energy $F(E_R)$, as well as detector atomic mass
number $A$, which would both typically influence the detection
rates. Instead, we restrict our analysis on the detectability of DM to
the astrophysical quantities, $\rho_0$ and $g(\vmin)$. We also 
restrict our analysis to the range of $m_\chi=5-10~\GeV$ and $\sigma_\chi=10^{-40}~{\rm
  cm^2}$ throughout the rest of this section. These benchmark rates can simply be
scaled for different values of $m_\chi$, $\sigma_\chi$, $A$, and
$F(E_R)$ as dictated by detections and individual experiments.

In the following subsections, we examine and describe the results from the individual models in detail, pointing out the physical mechanisms responsible for the observed rates.

\subsubsection{Fiducial Dynamical NFW Model} \label{subsubsec:fidmodel}

Calculating the detection rates hinges on accurately determining the
product of $\rho_0$ and $g(\vmin)$. We have presented the magnitude of
the density variations from spherical in
sections~\ref{subsubsec:dmdens} and \ref{subsubsec:wake}. We find that
the in-plane value can be increased by 50\% relative to the spherical
average, while the azimuthal variations can add up to an additional
40\%.  The deviation from an isotropic velocity distribution was
discussed in Section~\ref{subsubsec:dmvels}; both the shift of the
peak and the modification of the high-speed tail changes the DM
detectability.  At low $\vmin$, the increase in non-spherical density
dominates the signal, while at high $\vmin$, the deviation from an
isotropic velocity distribution significantly enhances the signal.

In the upper panel of Figure~\ref{fig:geevee}, we plot $g(\vmin)$ as a
function of $\vmin$. The distribution at the solar position as
calculated from the simulation is shown as a solid black
line. Uncertainties in the azimuthal position of the Sun are
represented by the
dark gray shaded region, while uncertainties as a result of the combination of
both the 
radial and azimuthal uncertainty are represented as the light gray
shaded region. The radial
uncertainty of the solar position relative to the length of the bar
causes significant deviations, with a trend to lower $g(\vmin)$ as the
radius increases. The azimuthal uncertainty is also significant, even
for a single choice of the solar radius. The value of $g(\vmin)$ increases
as the angle to the bar decreases, peaking when just slightly lagging the
bar (at a position angle of $-10^\circ$). The uncertainty increases
greatly at $\vmin>550\kms$, the result of a strong velocity
distribution component, as illustrated by the uncertainty in the speed
distribution based on choice of location (see Figure~\ref{fig:solvel}).

In Figure \ref{fig:geevee}, the dot-dashed and dashed black lines depict
$g(\vmin)$ for the pristine NFW profile and the adiabatically
contracted NFW model, respectively. These will facilitate
comparisons with all DM-detection experiments and can help to isolate
the effect of the dark disk and the shadow bar.  We analyze this further
in section~\ref{subsubsec:initnfw}. The SHM model is shown as the
solid gray line, which will be discussed further in
section~\ref{subsubsec:shm}. 

\begin{figure} 
\centering 
\includegraphics[width=3.3in]{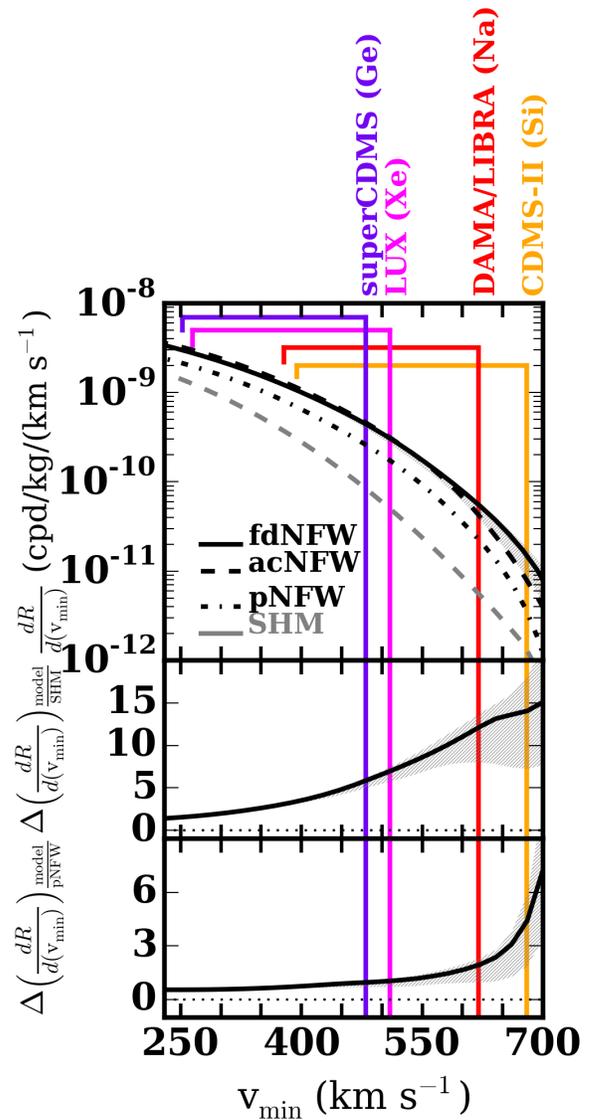} 
\caption{Upper panel: $dR/d(\vmin)$ as a function of $\vmin$ for
  various halo models. Line styles are the same as in
  Figure~\ref{fig:geevee}. The shaded region around the fiducial NFW model
  (black line) represents the total positional uncertainty effects on
  both density and the velocity distribution. Middle panel: detectability relative to the
  standard halo model, $(dR/d(\vmin)_{\rm model} - dR/d(\vmin)_{\rm
    SHM})/(dR/d(\vmin)_{\rm SHM})$. The shaded region again reflects
  the total uncertainty from both density and velocity distributions. Bottom panel: detectability
  relative to the pristine NFW model, $(dR/d(\vmin)_{\rm mode} -
  dR/d(\vmin)_{\rm pNFW})/(dR/d(\vmin)_{\rm pNFW})$. The shaded
  region is the same as in the middle and upper panels. The vertical 
  lines indicate the reported sensitivity limits for several direct
  detection experiments at $m_\chi=5$ GeV. The experiments are labeled
  above the figure, with the target nuclei listed in
  parentheses. Experiments are discussed further in
  Section~\ref{subsec:experiments}. Each experiment also has a
  horizontal line spanning $m_\chi=10$ GeV (left) to $m_\chi=5$ GeV
  (right, connecting to the vertical line) to
  demonstrate how the $\vmin$ threshold would change as a function of
  WIMP mass. \label{fig:detectrates}}
\end{figure}

In Figure~\ref{fig:detectrates}, we present the detectability of DM for
the simulations presented in \cite{petersen15}. We use
equation~\ref{eq:detect} to calculate $dR/d(\vmin)$ as a function of
$\vmin$.  The absolute detection rates are scalable for different
nuclear and DM parameters, but the dominant shape of the curve is given by
$\rho_0$ and $g(\vmin)$. The curves are plotted using the same scheme as
in Figure~\ref{fig:geevee}. For ease of interpretation,
Figure~\ref{fig:detectrates} also has vertical lines indicating
experimental detection limits at $m_\chi=5$ GeV (as well as a horizontal
 line to indicate the $\vmin$ values as $m_\chi$ increases to 10 GeV,
 at the left edge of the line), discussed further in
section~\ref{subsec:experiments}.

While both $g(\vmin)$ and $\rho_0$ increase with
$\Delta T_{\rm se}$, the corresponding scaling change required to hold
the bar radius fixed (as discussed in Section~\ref{subsubsec:dynunits})
leads to a decrease in $\rho_0$ with increasing $\Delta T_{\rm
  se}$. Thus, the overall results for $dR/\vmin$ are not strongly
dependent on $\Delta T_{\rm se}$, despite the dependence of the
individual factors on $\Delta T_{\rm se}$. An accurate age and
formation history for the MW bar will lead to a more precise
prediction.

In general, an enhancement relative to the SHM qualitatively means that
exclusion in $m_\chi-\sigma_\chi$ space becomes more
stringent. However, the subtleties of the shape of $g(v_{\rm min})$
(see Figure \ref{fig:geevee}) make placing experiments on the $m_\chi-\sigma_\chi$
plane difficult. From equation~\ref{eq:detect}, we see that the variation of
$g_{\rm NFW}/g_{\rm SHM}$ with $\vmin$ implies that for a fixed number
of detections $dR/d\vmin$, $\sigma_\chi/m_\chi$ will have an
inverse dependence on $\vmin$ for low $m_\chi$.

\subsubsection{Idealized NFW models} 
\label{subsubsec:initnfw}

We compare the fiducial dynamical NFW model to idealized NFW
models. First, we discuss the results as compared to the pristine NFW
profile, then discuss the adiabatically contracted NFW model.

The pristine NFW profile (eq.~\ref{eq:nfw}) is already demonstrably
different from the SHM, in both $\rho_0$ and $f(v)$ or $g(\vmin)$
(this is discussed further in section \ref{subsubsec:shm}). To
understand the effect that the dark disk and shadow bar have on the
detectability of DM, we compare to the pristine NFW profile rather than to
the SHM.  The relative enhancement factors for the fiducial dynamical and
adiabatic contraction NFW models are depicted in the lower panels of
Figures~\ref{fig:geevee} and \ref{fig:detectrates}.  We find that the
pristine NFW profile largely describes $g(\vmin)$ below
$\vmin=400~\kms$ and to within 50\% up to $\vmin=550~\kms$, above which the fiducial model turns up
sharply and the adiabatically contracted model turns up slightly. The
sharp upturn of the fiducial model owes to the
response of the DM particles to the bar. However, the bottom panel of
Figure~\ref{fig:detectrates} shows that the density increase enhances
the detectability relative to the pristine NFW profile. When the fiducial model is compared to the adiabatically contracted model (the
solid gray line in Figure~\ref{fig:geevee}), we see that the effect is
roughly the same below $\vmin=550\kms$, implying that the
variation owes primarily to the dark disk, an effect present in both
simulations. In Figure~\ref{fig:detectrates}, the fiducial and
adiabatically contracted models are largely the same below
$\vmin=550\kms$. Above $\vmin=550\kms$, the fiducial and adiabatically
contracted models deviate, indicating that the effect results from the
wake. The $\vmin$ value above which the
adiabatically contracted model and the fiducial model diverge varies weakly
with the secular evolution time, $\Delta T_{\rm se}$. As $\Delta
T_{\rm se}$ increases, the point of deviation moves 
to lower $\vmin$. As the in-plane DM density is approximately 10\%
larger in the fiducial model, an offset exists $dR/d\vmin$
between the two models, but the $\vmin$ value where the two models
begin to deviate is the same as in Figure~\ref{fig:detectrates}.

 In addition, the range in $g_{\rm
  NFW}/g_{\rm SHM}$ owing to solar position uncertainties increases with $\vmin$,
indicating that predicting detection rates at high $\vmin$ may be
particularly difficult until the MW bar parameters are more precisely
constrained.  As discussed in Section~\ref{subsubsec:fidmodel}, the range in
these ratios owe solely to the fact that the uncertainty in the angle and radius of the
solar position relative to the bar are
large.  Constraining the radial and angular position of the Sun relative to the MW bar,
as well as the fundamental parameters of the MW bar, is crucial to
accurately predicting the DD rates.

\subsubsection{Cored and rotating NFW models} 
\label{subsubsec:ohalos}

In section~\ref{subsec:methods}, we discussed the selection of the
fiducial NFW model in a cosmological context, noting that other halo
models could also meet the cosmological criteria. In this section, we
describe variations that result from changing this choice. Details of the
supporting simulations are presented in \cite{petersen15}.

For the cored NFW profile, the speed distribution peaks at even higher speeds
than the fiducial NFW model presented in Figure~\ref{fig:solvel}, up to
$+80~\kms$. The broadness of this distribution leads to an even larger
detectability compared to the SHM, up to a factor of 25 at
$\vmin>650~\kms$. Interestingly, the radial velocity peak is not
significantly shifted (in contrast to $\delta v_r=-30~\kms$ for the
fiducial NFW
model). This suggests that the shift in the speed distribution owes
to an increase in the azimuthal velocity as a result of rapid
angular momentum transfer during bar formation in this simulation
(see \cite{petersen15}).

The rotating halos demonstrate similar radial velocity shifts to their
nonrotating counterparts.  Specifically, the fiducial and rotating
NFW halos both peak at smaller radial velocity than their cored counterparts.  However, the azimuthal velocity peaks
for both rotating models are shifted to significantly higher values,
$>$100$\kms$ for some possible solar positions. This
shift owes to additional angular momentum transfer that creates an
even larger density in the galactic plane, which can begin to rotate like the stellar disk. The speed distributions for the rotating
models demonstrate a clear shoulder where the dark disk contribution
provides an excess signal near $v=450~\kms$, similar to
the findings in \cite{purcell12} for a particularly strong dark
disk. Thus, the rotating models are the easiest to detect, adding an
additional 50\% enhancement in $g(\vmin)$ over their non-rotating
counterparts (see Figure~\ref{fig:geevee}).

While each model is cosmologically consistent, rotating and
non-rotating models may represent qualitatively different initial
conditions in a cosmological setting. For instance, if the presently
observed stellar bar is not the first bar to have formed in the MW,
the DM halo may be imprinted with a relic response to a bar or other
strong bisymmetric structure (e.g. spiral arms) from the past that
have decayed or dissipated since those early times.  Further study of the history of the
MW bar and the stellar populations in the disk may help determine
the formation time of the MW bar and the likelihood that either a
previous bar existed or that the current bar had
significantly different parameters in the past. A triggered bar may
begin as a longer structure and subsequently shrink--in such a
scenario, the non-isoptropies generated by such an ancient bar may remain in the halo, adding
further substructure that is not present in our isotropic initial conditions.

\section{Discussion} 
\label{sec:discussion}

We begin this section with a discussion of our results in the context of the
literature (section \ref{subsec:litmod}), then discuss the implications of our fiducial model for DD
experiments, first as absolute sensitivities in section
\ref{subsec:experiments}, then for experiments that are sensitive to annual
modulations in section \ref{subsec:annualmod}.

\subsection{Literature halo models} 
\label{subsec:litmod}

The DD literature is largely dominated by use of the SHM. To connect with those results, we analyze our models and the results presented in \ref{subsec:detectionrates} and compare to the SHM. In the absence of measurable
density and velocity profiles for the MW DM halo, the SHM has been
used as a benchmark. However, extensive reports exist in the
literature (e.g. \cite{kuhlen10,purcell12,pillepich14}) regarding the
inaccuracy of this model compared with cosmological simulations,
though recently, studies have claimed that the SHM may be a viable
model \cite{bozorgnia16,kelso16}. In these studies, empirical halo models have been used to constrain
the parameter space for dark matter properties in the
$m_\chi-\sigma_\chi$ plane. We qualitatively discuss the
results of those works in section \ref{subsubsec:modelcomp} and attempt to reconcile the results using the
physical explanations presented in section \ref{subsec:detectionrates}.

\subsubsection{The Standard Halo Model} \label{subsubsec:shm}

The SHM has
a density profile of $\rho\propto r^{-2}$ to satisfy the requirement
of a flat rotation curve at the solar circle, normalized such that
$\rho_0=0.3~\GeV$ is the density at the solar circle, with an isotropic velocity distribution given by a
MB distribution
\begin{equation}
f(v) = 4\pi v^2 \exp\left(-\frac{v^2}{2\sigma^2}\right)
\end{equation}
with $\sigma = v_{\rm LSR}/\sqrt{2}$ and $v_{\rm LSR} = 218~{\rm
  km~s^{-1}}$. Because the MB
distribution has infinite tails, the SHM typically includes a
truncation for the galactic escape speed, either by using an error
function or by subtracting a MB distribution with a
velocity $v_{\rm esc}$.  Several studies have investigated the
galactic escape speed using stellar kinematics, with findings ranging
from $v_{\rm esc} = 533^{+54}_{{-41}}$ \cite{piffl14} to $v_{\rm esc}
= 544^{+64}_{{−46}}$ \cite{smith07} to $v_{\rm esc} = 613$
\cite{piffl14a}.

We will compare the SHM to the fiducial dynamical NFW model by choosing $v_{\rm esc}$ to
be the
highest velocity particle in the simulation, and note the effect of a
lower galactic escape speed where relevant (see \cite{lavalle15} for
an investigation of the explicit effects of escape speed
choice). Conversely, some simulation particles will have speeds higher
than the nominal escape velocity. Although these may be transient
particles that are not bound to the DM halo, these particles
will still contribute to the signal. This is likely for the real MW as well
and thus motivates our choice to depart from literature choices of
$v_{\rm esc}$ for the purpose of this comparison, and instead apply
our own empirical $v_{\rm esc}$ to perform the analysis. This
may be a large source of the disagreement between these findings and
other works.

The middle panels of Figures~\ref{fig:geevee} and
\ref{fig:detectrates} demonstrate the strong detection enhancement for the
fiducial NFW profile relative to the SHM. Figure~\ref{fig:geevee}
presents the effect of the velocity structure alone.
Figure~\ref{fig:detectrates} compares the computations of
Equation~\ref{eq:detect} and
describes the effects of both the velocity and the density; i.e., the total
effects of the more realistic NFW halo model. Owing to the
broadening in the model speed distribution when compared to the SHM,
$g_{\vmin}$ is enhanced for all $\vmin$ and increases with increasing
$\vmin$. Figure~\ref{fig:geevee} shows that the velocity distribution function
alone yields a factor of four increase at high $\vmin$, steadily
increasing for all $\vmin$. Figure~\ref{fig:detectrates} shows that the estimates for
DM detection rates may be 20 times larger than the SHM estimates for
some experiments as a result of the strong enhancements of $g(\vmin)$ and
$\rho_0$.

Best-fit MB distributions will indeed overpredict the tail of the
velocity distribution, consistent with findings in the literature
\cite{sloane16}. However, the SHM is not measured as a best-fit MB
distribution, but rather a specific evolution-dependent distribution
as described in section \ref{subsubsec:fidmodel}. While using a
parameterization for the velocity distribution that includes $v_{\rm
  esc}$ may be tempting to ease the tensions between lighter and heavier
nuclei experiments, our results indicate that there is little
dynamical reason to expect a strong dependence of the shape of the
velocity distribution on $v_{\rm esc}$. Additionally, the tensions
between ligher and heavier nuclei experiments cannot be resolved
with our models.

\subsubsection{Simulation-based models}\label{subsubsec:modelcomp}

We first discuss the reported simulated DM density and velocity distributions in the literature before making a
direct comparison to our work. We then discuss potential dynamical
reasons for the differences.

In the absence of strong constraints on the DM density at the solar
circle, simulations which attempt to match various other parameters to
define a `MW-like' galaxy have a variety of DM densities at the solar
circle. In particular, while some studies explicitly discuss the
presence of a dark disk \cite{purcell12,pillepich14}, others find no
evidence for a dark disk \cite{kelso16}, and others still find a dark disk in
some simulations but not others
\cite{bozorgnia16,sloane16}. No previously reported simulations
attempt to characterize the dependence of DM density on azimuth.

In addition to the variations in DM density, the reported velocity
distributions of the simulations vary considerably. Generally, studies seek to explain the speed distribution through a
parameterization at least reminiscent of the MB distribution. Upon
inspection of various velocity components in this work (see section \ref{subsubsec:dmvels}), it is not
clear why a MB-derived one-dimensional speed distribution should be
expected. In examining literature examples, each dimension of the velocity
distribution appears to depart from Gaussians.

An attempt to find an empirical form to describe a halo velocity distribution function
led to the result of \cite{mao13}, which parametrizes the speed
distribution as a function of the escape velocity and a parameter $p$
that controls the steepness of the tail of the distribution, such that
the tail approaches an exponential distribution at low velocities instead of
a Gaussian. \cite{pillepich14} find that the speed distribution parameterization of \cite{mao13} fits their
empirical velocity distributions better than the SHM. 

In addition to the fully self-consistent simulations in \cite{mao13}
and \cite{pillepich14},
\cite{fornasa14} constructed a model that
allowed for an anisotropic velocity distribution in the DM of the MW,
and used an extended Eddington inversion formalism to calculate the
distribution function including the separate mass components of the MW (stellar disk,
bar, bulge, interstellar medium, DM halo). Relaxing the assumption of
isotropy by including different mass components increased the
parameter space of $f(v)$, including a factor of 2 change in the
high-velocity tail. These results are consistent with the
phenomenological $N$-body simulation parametrization of
\cite{mao13}. 

In light of our findings presented in this paper (section \ref{subsec:features}), we discuss the
compatibilities of our results with the simulations discussed above. The largest difference between previous empirical halos and our work
is the inclusion of the bar dynamics and its resulting DM
response. In particular, several papers with which we compare results
analyze galaxies with no apparent bar
\cite{vogelsberger09,kuhlen12,pillepich14, bozorgnia16, kelso16, sloane16}. Previous studies have also
focused on contributions from a dark
disk \cite{read08,read09}, tidal streams
\cite{vogelsberger09,kuhlen12,ohare14}, and debris flows
\cite{kuhlen12,lisanti12}.  

Regardless of the included prescriptions for various astrophysical
processes or included components, simulations must adequately describe gravity and address
the findings presented here (the dark disk and kinematic structure
resulting from both the disk and stellar bar).  Thus, it is difficult to reconcile simulations that do not
observe an in-plane overdensity
\cite{bozorgnia16,kelso16}, or those with little in-plane overdensity
\cite{pillepich14} with this work (section \ref{subsubsec:dmdens} and
\ref{subsubsec:wake}) and the associated dynamical
results in \cite{petersen15}. 

The dearth of dark disk material may be due to
merger history (as has been claimed), though the simulations of \cite{sloane16} appear to
show that models of the MW that have undergone recent quiescent
periods still support our findings regarding the influence of bar-driven
dynamics.  We conjecture that \cite{kelso16} and
   other simulations are inhibiting the formation of a dark disk as a
   natural response to the stellar disk regardless of the merger
   history (both simulations discussed in \cite{kelso16} have a
   relatively quiescent history). Possible causes include the
   initial temperature of the halo (as measured in velocity
   dispersion), over-heating of the stellar (and therefore dark) disk,
   and insufficient potential and phase space
   resolution. \cite{bozorgnia16} does not provide enough information
   on merger history for us to make even a qualitative assessment of their dark
   disks (or lack thereof).

We have demonstrated in \cite{petersen15} that in sufficiently
accurate simulations
secular processes change both the ratio of the radial to azimuthal
action, which manifests as a change in orbital eccentricity, and
induces a net rotation.  Thus, DM particles secularly evolve into dark
disk orbits.  As described in section~\ref{subsubsec:dmvels}, these
effects are both at play in the velocity structure presented here. Both \cite{bozorgnia16} and \cite{kelso16} report bulk
rotation ($\delta v_\theta\approx20~\kms$) in their DM halo models, albeit at a smaller $\delta v_\theta$ than
reported in our simulations. As shown in Figure \ref{fig:correlatedvel}, the $v_r-v_\theta$
relationship is altered by the presence of the quadrupole wake, which
results from the stellar+shadow bar. The deviations may be below the
sensitivity threshold of other simulations, in particular those that
cannot probe the $v_r-v_\theta$ plane a function $\Delta \theta_{\rm
  bar}$. If the numerical 
sensitivity does not allow for a characterization of these deviations,
then we would expect them to recover a MB distribution, consistent with the
findings of \cite{bozorgnia16,kelso16,sloane16}. 

In previous works, the limits and detection regions imposed by DD
experiments are primarily affected by the density distribution at the
high-mass end ($m_\chi>10$ GeV), while both the velocity distribution
and density of the self-consistent models affect the low-mass end
($m_\chi<10$ GeV). As $\vmin$ increases, the $\sigma_\chi-m_\chi$ parameter space covered is
particularly sensitive to DM halo model choice. Above $m_\chi=20$ GeV,
the velocity differences are less pronounced, but the $\rho_0$
determination is still crucial for placing accurate limits. The parametrization presented in \cite{mao13} allows for a steeper fall-off in
the speed distribution, which may alleviate some of the tension
between DD experiments (see Section~\ref{subsec:experiments}), though
this has not been functionally demonstrated \cite{pillepich14, bozorgnia16, sloane16}. The next
section discusses the effect of our models on the interpretation of DD
experiments. 

The results from \cite{pillepich14} and \cite{fornasa14}
are generically consistent with results for the adiabatically
contracted NFW 
model, but fail to match the secular evolution caused by the bar, an effect we have shown is significant to the
prediction of the DD rates. In both our work
and \cite{purcell12}, the inclusion of the stellar disk
potential increases $g(\vmin)$ by broadening the speed distribution in
the plane. The overall DM detection rates presented
here are qualitatively similar to those in 
\cite{purcell12}, but for different physical reasons. In our model, the
uncertainty in the solar position, which may contribute a factor of
two to the detection rates,
is significantly larger than the variation between the models in
\cite{purcell12} (approximately 40\% at the largest). As \cite{purcell12} seeks to model the effect of the Sagittarius dwarf
(a satellite of the Milky Way presently having strong interactions
with the disk), their $<$40\% result, when compared to our $>$100\% result, suggests that
the stellar+shadow bar is a significantly larger effect than the
Sagittarius dwarf for all realistic assumptions about the stellar bar
and the
Sagittarius dwarf.

We note that these cosmologically-based
studies (\cite{kuhlen10,purcell12,pillepich14,bozorgnia16,kelso16,sloane16}) do have advantages when compared to the models presented here, namely added realism from the growth
of the stellar disk over time, as well as the presence of substructure
in a DM halo that evolves self-consistently. We intend to address
the generic dynamical effects of these phenomena in future
work. Regardless, the dynamical findings
that manifest as detectable signals in this work are
bolstered by theoretical predictions (e.g. \cite{weinberg85}). Further, it is difficult to see how other dissipational-component specific processes
(e.g. star formation, feedback) would preferentially affect the halo;
we therefore expect the results presented in this paper to be generic.

\subsection{Implications for Direct Detection Experiments} 
\label{subsec:experiments}

Clearly, no simulations can yet make robust
predictions for absolute DD rates in the MW. Qualitatively, the
increased detection rates observed in simulations relative to the SHM is a
boon to DD experiments. Perhaps more importantly, to accurately
interpret DD experiment results, and when comparing different DD experiments, the speed distribution is the
largest uncertainty (see the discussion in Section
\ref{subsec:detectionrates}). Because the speed distribution is composed of
the three components of the velocity, changes to the Gaussian nature
of any of these distributions will result in a non-MB velocity
distribution. 

In Figure~\ref{fig:detectrates}, the approximate sensitivities to
$m_\chi=5\GeV$ DM are plotted as vertical lines to
illustrate the potential cumulative effect the dark disk, density
wake, and kinematic wake can have for various
experiments (see
\cite{aalseth13,bernabei10,agnese13a,angloher16,aprile10} for
sensitivity determinations, where $E_R$ has been translated to
$\vmin$ as in Section~\ref{subsec:detectionrates}).  Each experiment
has been able to place limits on $\sigma_\chi$ and $m_\chi$, with the
earlier generation CDMS-Si experiment \cite{agnese13a} finding three possible events that
make the most likely model for a DM particle $m_\chi=8.6 \GeV$ and
$\sigma_\chi = 1.9\times10^{-41} ~{\rm cm^2}$, consistent with the CoGeNT
results \cite{aalseth13}, as well as the DAMA (Na) results
\cite{bernabei14}. We also plot horizontal lines connecting the
vertical line at $m_\chi=5$ GeV (right extent) to a limit at
$m_\chi=10$ GeV (left extent) as a function of
$\vmin$ to demonstrate the different values of $dR/d\vmin$ each
experiment would reasonably expect to observe.

Recently, tensions between different experiments, notably the LUX,
XENON100, superCDMS
and CDMS-Si experiments have been reported. \cite{frandsen13} appears to
find that varying astrophysical parameters cannot explain the observed
CDMS-Si and XENON100 tension, which our findings support.  As discussed in
section~\ref{subsubsec:shm}, the dependence of $g_{\rm NFW}/g_{\rm
  SHM}$ on $\vmin$ suggests that experiments with significantly
different $\vmin$ thresholds will be up to 10 times discrepant in
their detection rates for realistic galaxy models when compared to the
SHM at $m_\chi=5$ GeV.  Of course, the experiments sensitive to the lowest energy
thresholds still have the largest absolute values of $g(\vmin)$, but
the relative ability to detect $m_\chi=5~\GeV$ DM for experiments
with higher energy thresholds is significantly enhanced (middle panel
of Figure~\ref{fig:geevee}).  Specifically, the detection rates for
the CDMS-Si experiment increase by a factor of $>$15 (4) at $m_\chi=5$
GeV ($m_\chi=10$
GeV) while for the
LUX the detection rates increase by a factor of 7 (2) at $m_\chi=5$
GeV ($m_\chi=10$
GeV). Thus if CDMS-Si had set the same limit as LUX using the SHM as
the halo model, the limit of CDMS-Si would actually be twice as
sensitive if one used a more realistic halo model. However, the low
energy threshold of LUX (1.1 keV, \cite{akerib16}) still allows LUX to
set the more stringent limit.

\subsection{Implications for Annual Modulation Signals}
\label{subsec:annualmod}

\begin{figure} \centering 
\includegraphics[width=3.2in]{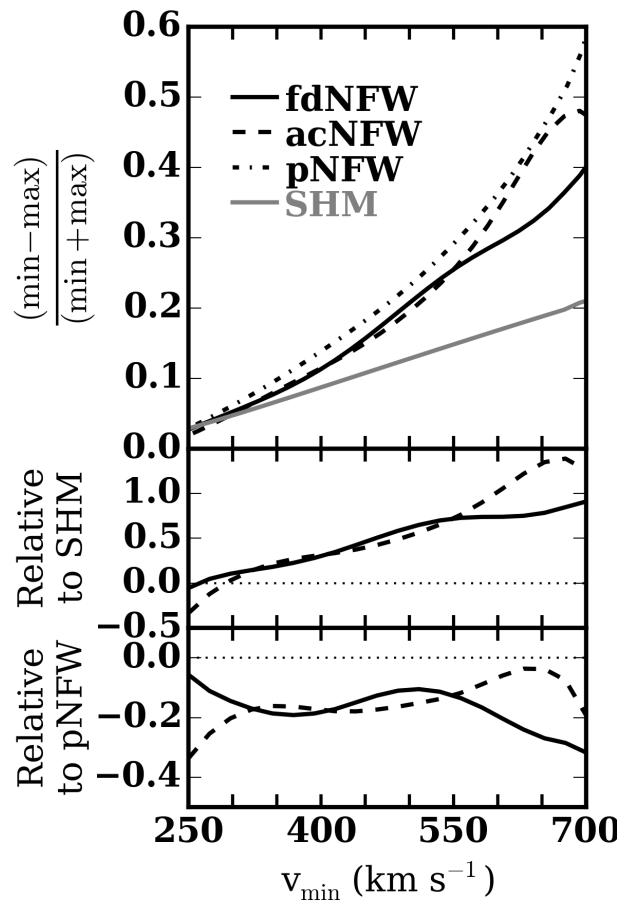} 
\caption{Upper panel: Annual modulation fraction, $(R_{\rm max} -
  R_{\rm min})/(R_{\rm max} + R_{\rm min})$, as a function of $v_{\rm
    min}$. The models are shown following the same convention as in
  Figures~\ref{fig:geevee} and \ref{fig:detectrates}. Middle panel:
  relative enhancement factor for the fiducial dynamical NFW model and the
  adiabatically contracted NFW model, compared to the SHM. Bottom panel: relative enhancement factor for the
 fiducial model and the adiabatically contracted model, compared to
  the pristine NFW profile. \label{fig:annmod}}
\end{figure}

For an isotropic DM distribution velocity distribution in the LSR
frame, an annual modulation of the DM signal will arise from the
oscillation of the Earth's azimuthal velocity ($V_\oplus$) between
its minimum and maximum values relative to the DM halo.  This modulation has
been fit by a sinusoid that peaks at the day of highest azimuthal
velocity (e.g., \cite{pillepich14}). In our dynamical model, two effects are at play: the
modulation will be affected by asymmetries in the velocity centroid
and the shape of the velocity
distribution with respect to the LSR (as described in section
\ref{subsubsec:dmvels} and illustrated in Figures \ref{fig:solvel} and
\ref{fig:correlatedvel}).

In the upper panel of Figure~\ref{fig:annmod}, we plot the amplitude
of the annual modulation as the difference between the minimum and
maximum detection rates during a year, $(R_{\rm max} - R_{\rm
  min})/(R_{\rm max} + R_{\rm min})$ as a function of $v_{\rm
  min}$. The annual modulation amplitude increases in all models with increasing
$v_{\rm min}$ but, owing to adiabatic contraction, the modulation in
both the adiabatically contracted model and the fiducial model are
highly enhanced, particularly at high $v_{\rm min}$. The differences
between the adiabatically contracted model and the fiducial dynamical
simulation are caused by the stellar and shadow bar.

In the middle panel of Figure~\ref{fig:annmod}, we compare the
fiducial and adiabatically contracted models to the SHM. Compared
to the SHM, both the adiabatically contracted model and the
fiducial model are enhanced for $v_{\rm min}>300~\kms$, of
interest to most detection experiments (also pointed out in
\cite{purcell12,pillepich14}). In the bottom panel of
Figure~\ref{fig:annmod}, we compare the fiducial and adiabatically
contracted NFW models to the pristine NFW profile. Here, we see an
opposite effect to the comparisons to the SHM: the annual modulation signal is
decreased.  

Clearly, dynamical evolution affects the
annual modulation predictions. We now focus on the comparison between the fiducial and
adiabatically contracted model to isolate the effect of the stellar+shadow bar.  The velocity ellipsoid of the fiducial model is
 isotropic and skewed to lower radial and higher tangential
velocities, in contrast to the adiabatically contracted model (and the
SHM), as shown in section~\ref{subsubsec:dmvels} and
Figure~\ref{fig:correlatedvel}.

A comparison of the fiducial and
adiabatically contracted models illustrate the effect of the
anisotropic velocity ellipsoid on the annual modulation. For
$v_{\rm min}<300~\kms$, the amplitude of the fiducial model is
enhanced relative to the adiabatically contracted model, while for $v_{\rm
  min}>550~\kms$, the adiabatically contracted model is enhanced relative
to the fiducial model. In \cite{purcell12}, the Sagittarius
stream DM material is out of phase with the annual modulation signal
(the stream originates from galactic north). We find that the annual
modulation signal in their simulations will closely match the result
of our adiabatically contracted model, due to the contribution of the
dark disk.

However, in the presence of the bar feature, differences arise.  We
find that the tail of the speed distribution is dominated by orbits
tangential to the LSR motion, but owing to the difference between the
expected annual modulation velocity vector from an isotropic
distribution and the solar velocity vector (see
Figure~\ref{fig:correlatedvel}), the effect is lessened as some of the
DM co-rotates. However, at low velocities, the radial velocity peak
being centered at $v_{r}<0$ contributes some signal relative to the
adiabatically contracted model.

\cite{freese13} provides an overview of the prospect for annual
modulation given the status of DD experiments; we point out here that
while the overall amplitude of the annual modulation detection signal in our NFW model increases relative to
the SHM, the effect of the stellar+shadow bar reduces the effect at
high velocities, increasing it at low velocities. As the absolute
scaling of the amplitude depends on the location of the peak of the
speed distribution relative to the annual modulation velocity
variation, we cannot definitively say that the annual modulation
signal will be increased.  Nonetheless, the trends in the current
experimental data are broadly consistent with the isolated effects of
the shadow bar provided by the fiducial dynamical NFW and adiabatic
contraction NFW models: experiments with low energy thresholds have
reported possible annual modulation signals, and high energy
threshold experiments have not.

\section{Conclusions} 
\label{sec:conclusion}

The major results of the paper are as follows:

\begin{enumerate}
\item The density of the DM halo at the solar position varies depending on the Earth's location relative to the stellar bar. Smaller
  angles relative to the bar as well as a smaller ratio of $R_{\odot}/R_{\rm bar}$ can increase the density relative to a spherical
  distribution by a factor of 2.

\item The DM velocity profile is reshaped by the stellar+shadow
  bar. The characteristic quadrupole wake in the DM that forms as a
  response to the stellar bar lags the bar in velocity and, therefore,
  enhances the detectability of DM when compared to the SHM
  (adiabatically contracted NFWmodel) by a
  factor of 3.5 (2) at $\vmin=300~\kms$. At $\vmin=650~\kms$,
  detectability relative to the SHM is increased by a factor of 10,
  and up to a factor of 40 for a cored NFW halo model. Enhancements
  for initially rotating models are approximately equal to the
  respective non-rotating model (fiducial dynamical NFW and cored NFW).

\item A number of recent
  astrophysical models suggest the importance of the MW evolutionary
  history to modeling DM detection rates. As detectability depends on
  $\vmin$ (which is sensitive to the velocity distribution), and we
  have demonstrated effects on the velocity distribution from known
  features in the MW, experiments need to move beyond the SHM to
  compare with other experiments that have different energy
  thresholds.

\item Similarly, annual modulation in the DM signal will have
  different detectabilities compared to the SHM as a function of
  $\vmin$. The stellar+shadow bar, when compared to the adiabatically
  contracted model, {\it reduces} the annual modulation signal for
  experiments sensitive to high energy thresholds by approximately
  20\%, and {\it boosts} the annual
  modulation signal for experiments sensitive to low energy
  thresholds by approximately 20\%.

\item When compared to the SHM, we expect an enhancement in
  detectability and annual modulation.  We use an adiabatically
  contracted model that fixes the gravitational potential of the disk
  to calibrate the importance of dynamical evolution to the DM
  detection predictions.  For example, when we compare our fiducial
  dynamical NFW model to the
  adiabatically contracted NFW model at $\vmin=475~\kms$ (the nominal
  value for superCDMS at $m_\chi=5$ GeV), we expect an enhancement in
  detectability of 100\%, but an unchanged annual modulation signal.  This
  illustrates the influence of dynamical evolution.
\end{enumerate}

The results presented in this paper can be succinctly
  summarized as indicative that the expected rates of observation for DD
  experiments is strongly sensitive to realistic DM halos. Models that incorporate known physical
  processes can be used at a minimum to determine
  astrophysics-related constraints on DM $m_\chi$ and $\sigma_\chi$. While the literature now has no shortage of simulations touting
different halo velocity distributions, the field is still not able to accurately create a MW analogue that accounts
for evolutionary history. Acknowledging this fact, in this paper we
study the effects of simple dynamical models, implemented
through n-body simulations, on DD experiments. We
stress that the effects presented in this paper are generic results of the
gravitational interaction between the stellar disk and the DM
halo. The power in these inferences is a motivation for marrying DD
experiments with realistic astronomy. Astronomically realistic models will provide realistic constraints with more power to
discriminate between WIMP hypotheses.

The change relative to the SHM affect primarily lower $m_\chi$
values. This owes to the low $v_{\rm min}$ values implied by
$m_\chi>20$ GeV, allowing experiments to probe nearly the entire
$g(v_{\rm min})$ space. In contrast, if $m_\chi<10$ GeV, the
discrepancy between our fiducial model and the SHM will be large:
$v_{\rm min}$ is in the tail of the $g(v_{\rm min})$ distribution, where we have demonstrated $\left(\Delta g(v_{\rm min})\right)/g(v_{\rm min})$ changes rapidly.

The results presented here are by no means an exhaustive parameter
search, nor a best-fit MW model.  However, the MW is a disk galaxy
with a moderate bar.  The features induced in the DM distribution by
dynamical evolution in our simulations realistically represent those
expected in the MW and will obtain generally for any disk galaxy.  The
density enhancements and velocity asymmetries will have clear impacts
on the sensitivities of the various direct-detection experiments and are
likely to make the tensions between upper limits and tentative detections
stronger and more interesting.  Future iterations of direct detection
experiments, such as superCDMS (at SNOLAB) \cite{agnese15}, LUX-ZEPLIN
\cite{akerib15}, and XENON1T \cite{aprile14}, will build
upon the constraints from previous studies.  Halo models that
accurately account for known dynamical effects in the MW are necessary
for meaningful hypothesis testing.

Finally, directional detectors will enable a detailed study of the
kinematic signature at the solar position. Early efforts may be able
to detect a bias in the tangential and radial velocity peaks, as in
Figure~\ref{fig:correlatedvel}, which may even prove a
discriminating factor for determining the halo profile. This hints at
the possibility of DM astronomy in the future.

\section*{Acknowledgments}

The authors thank Eric Linder for thoughtful comments on a draft
version of this work.

\bibliography{PetersenM}

\end{document}